\documentclass[12pt]{article} 
\usepackage{graphicx}
\usepackage{amsmath}
\usepackage[top=30mm,bottom=30mm,left=25mm,right=25mm]{geometry}
\usepackage{setspace}

\newcommand{\bfig}[3]{\begin{figure}\centering\includegraphics[clip,width=10.0cm]{#1} \caption{#2} \label{#3} \end{figure}}

\newcommand{\afig}[3]{\begin{figure} \centering \includegraphics[width=17cm, clip]{#1} \caption{#2} \label{#3} \end{figure}}

\title{Robustness of a SiECAL used in Particle Flow Reconstruction}

\author{
Chihiro Kozakai$^1$, Shion Chen$^1$, Daniel Jeans$^1$, \\ Yoshio Kamiya$^2$, Sachio Komamiya$^1$\\ \\
$^1$Department of Physics, Graduate School of Science, \\ The University of Tokyo\\
$^2$International center for Elementary Particle Physics, \\ The University of Tokyo}

\date{2014/3/31}

\begin{document}

\maketitle

{\em Talk presented at the International Workshop on Future Linear Colliders (LCWS13),\\ Tokyo, Japan, 11-15 November 2013.}

\begin{abstract}

The physics program of future lepton colliders such as the ILC, will benefit from a jet energy resolution in the range 3-4\%.
The International Large Detector (ILD) reaches this goal over a large range of jet energies.
In this paper, we report on the dependence of the simulated ILD performance on various parameters of its silicon-tungsten ECAL.
We investigate the effects of dead areas in the silicon sensors, the thickness of the PCB at the heart of the detector,
and the robustness of its performance with respect to dead channels, noise, mis-calibration and cross-talk.

\end{abstract}

\newpage

\section{Introduction}

Many physics targets at the ILC require precise jet energy measurement \cite{TDR1}. 
A precision of 3-4 \% allows hadronic decays of W and Z bosons to be distinguished \cite{DBD}.
Particle Flow Algorithms (PFA) for jet energy reconstruction have been developed 
to meet this requirement \cite{PFA1}\cite{PFA2}. The basic idea of the PFA is that each particle should be measured with the most appropriate detectors. 
The energy of charged particles should be measured by the tracking system, while the calorimeters are used to measure the energy of neutral particles.
PFA relies on the ability to distinguish charged and neutral calorimeter clusters.

The ILD is optimized for PFA. A silicon tungsten electro-magnetic calorimeter (SiW ECAL) is proposed as one ECAL option for the ILD \cite{DBD}. 
The SiW ECAL is a sandwich-type calorimeter with layers of tungsten absorbers and silicon semiconductor detectors. The silicon detector layer 
has high granularity, with a segmentation of $5\times 5 \, \mathrm{mm}^2$ in the current design. The SiW ECAL is longitudinally segmented 
into 30 layers.

The performance of the SiW ECAL has been tested by both experiments and simulations \cite{DBD}. They showed that SiW ECAL has sufficient 
performance for the ILD. 
In this paper, optimization for the guard ring width and PCB thickness is discussed. Moreover, the robustness of the SiW ECAL for dead channels, 
noisy pixels, mis-calibration, and cross talk are studied. 

\section{Method of the simulation}
 
Events were simulated with the detailed full simulation of the ILD detector using Mokka \cite{Mokka}, a GEANT4-based \cite{GEANT4} 
simulation framework. The same detector model as used in the ILD Detailed Baseline Design \cite{DBD} was simulated. Here, detector 
parameters of guard ring width and PCB thickness are changed for the optimization studies. Event reconstruction was performed in the 
Marlin framework \cite{Marlin}. This includes digitization of the simulated detector signals (including a threshold cut at 0.5 MIP in each 
ECAL pixel), and reconstruction of events using PandoraPFA\cite{PFA2}. Effects for the robustness studies were added in the digitization step.

The performance of the 
ILD
was evaluated by the obtained jet energy resolution. The jet energy resolution is measured using a Z boson at rest decaying to 
two light quarks(u, d, or s). Since the energy resolution is worse in the overlapping region of barrel and end-caps and in the beam direction, 
only events with $|\cos \theta|<0.7$ are used, where $\theta$ is the angle between the initial quark direction and the beam. 
In addition, only 90\% of the events around the peak are used because the measured energy distribution has tail. 
The RMS of these 90\% of events, RMS90, is interpreted as a measure of the jet energy resolution of ILD. 

\section{Simulation results}

\subsection{Guard ring width}

The sensor is a matrix of Si PIN diodes. A guard ring is set on the edge of the silicon sensor. It prevents surface dark current, decreases noise 
and keeps dynamic range. This guard ring area is a less efficient area and making small guard ring with as good property as 
large one is difficult. In simulation, the guard ring is treated as an insensitive area.

First, it is investigated that how the guard ring effect can be seen. The barrel section of the ECAL consists of five modules in the beam 
direction, five columns of detector slabs per module, and two Si sensors per slab in the beam direction. 
Therefore, it is expected from the structure that particles which 
hit the ECAL perpendicularly will show a large guard ring effect, while particles which hit the ECAL with a large incident angle will be less affected. 

\afig{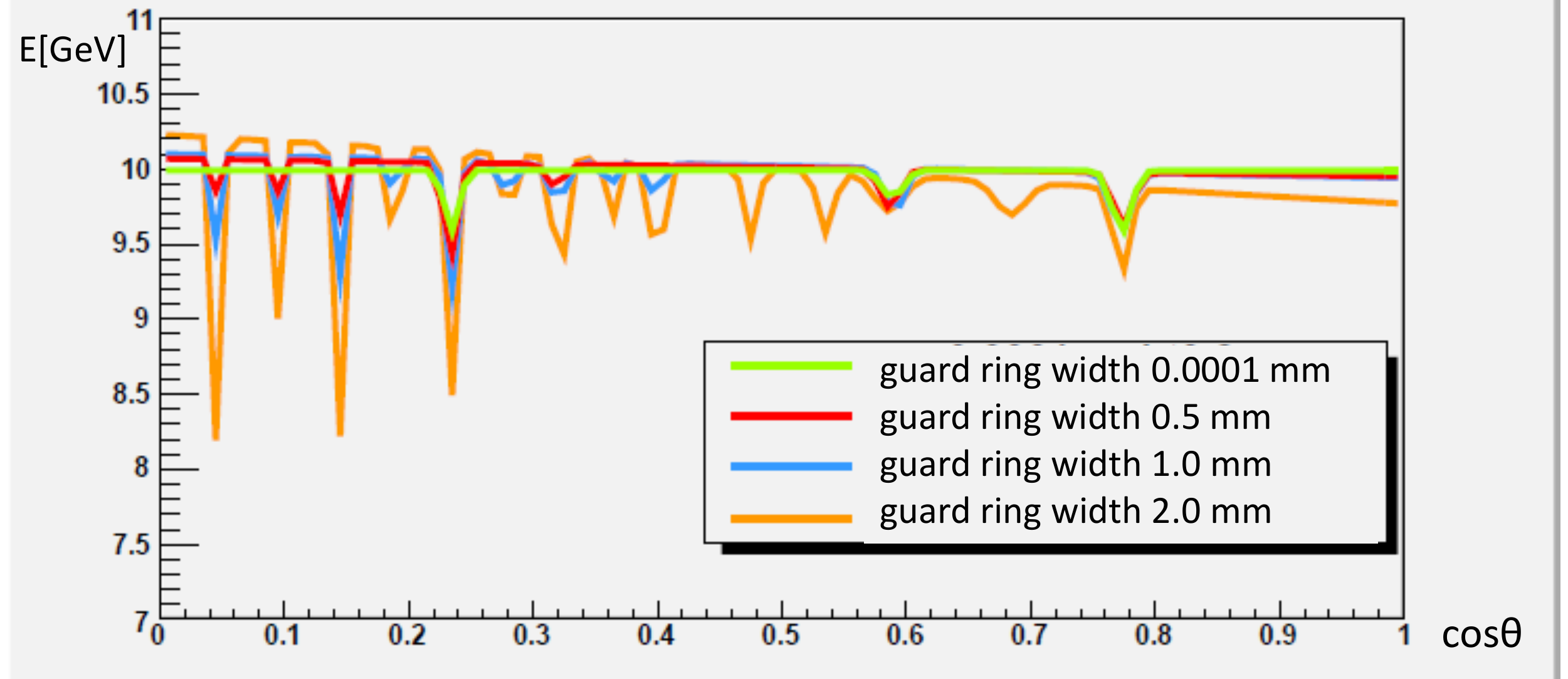}{The mean of the reconstructed energy function of 10 GeV photons, as a function of the cosine of the polar angle. }{energycorrection}

To see the effect, single 10 GeV photons were simulated. The plot of the mean reconstructed energy against the $\cos\theta$ was fitted by a 
linear function and some Gaussians. Fig.\ref{energycorrection} shows a result. Each Gaussian shaped dip corresponds to each guard ring, slab end, 
module end, or barrel and end-cap overlap region. As expected, there is large guard ring effect for large guard ring width. 
Especially, in middle cosine theta direction, function difference between different guard ring thickness is large. 

\afig{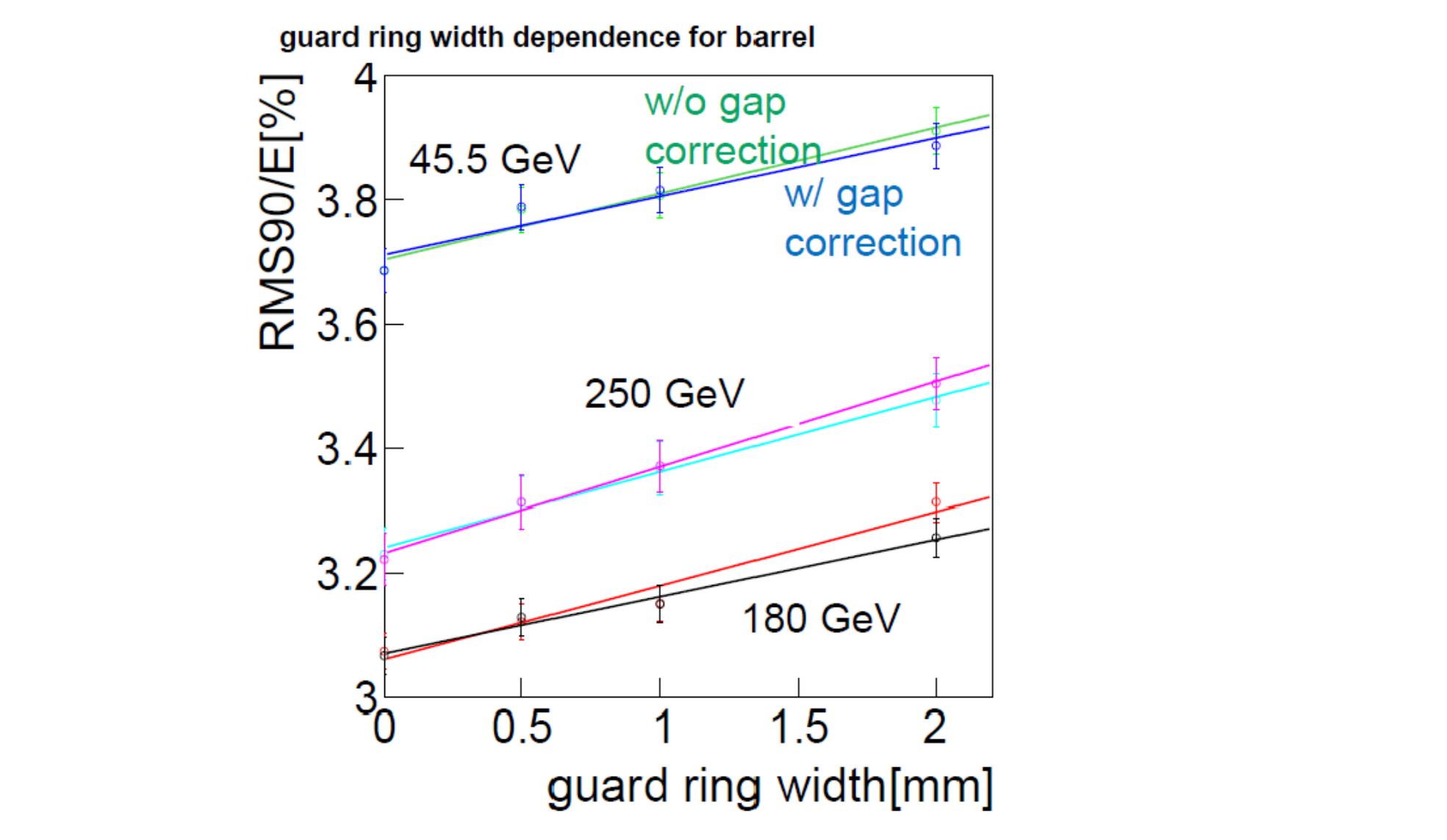}{The dependence of the jet energy resolution on the guard ring width.}{GRJER}

Figure \ref{GRJER} shows the dependence of the jet energy resolution on the guard ring width. The energy resolution degrades aproximately 
linearly with guard ring width. It increases by 6\% between 0 mm and 2 mm. Here, the result with gap correction for photon using a function 
shown in Fig.\ref{energycorrection} is also shown. However, this correction does not give significant improvement to the energy resolution. 
It is because the fraction of the photon is about 30\% and largely corrected photon is few.

\subsection{PCB thickness}

As there are many channels in the ECAL, PCB and electronics systems are put within each layer to serialize signals. The dead area due to 
readout cables is therefore reduced. A thick PCB may decrease the effective density of the ECAL, and therefore increase lateral shower size. 
Therefore, a thin PCB may be preferable. However, a thin and flat PCB is technologically difficult and expensive to produce. 

\bfig{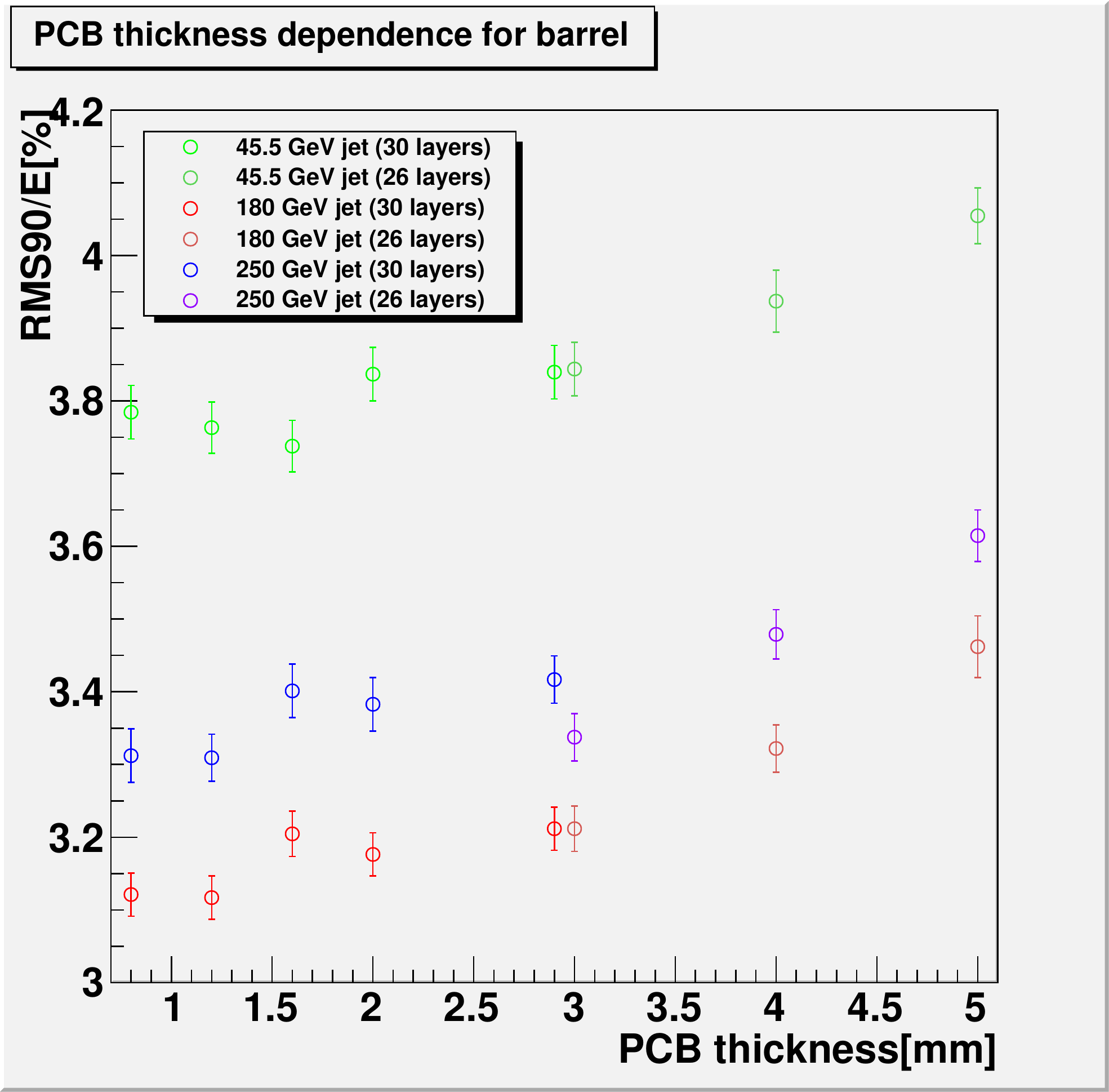}{The dependence of the jet energy resolution on PCB thickness, for different jet enegies.}{PCBJER}

Fig.\ref{PCBJER} shows the dependence of the jet energy resolution on the PCB thickness. The number of layers was reduced to 26 for PCB thickness 
above 3 mm 
for technical reasons of the detailed Mokka detector description.
It is clear from Fig.\ref{PCBJER} that this change in the number of layers does not affect the result. 
The degradation of jet energy resolution between 0.8 and 3mm is rather small, while for thicknesses above 3mm the degradation becomes
more significant.
The jet energy resolution increases by 5-8\% between 3 mm and 5 mm.

\subsection{Dead channels}

If a few \% of dead cell are allowed, the yield for Silicon sensor production may be increased and their cost reduced. 
To study the effect of such dead channels, a fraction of channels were randomly removed at the digitization step.
Each front end ASIC reads out 64 channels. The effect of broken readout ASICs (possible, for example, during constuction or operation) 
was studied in a similar way by randomly removing blocks of 8x8 channels.

\afig{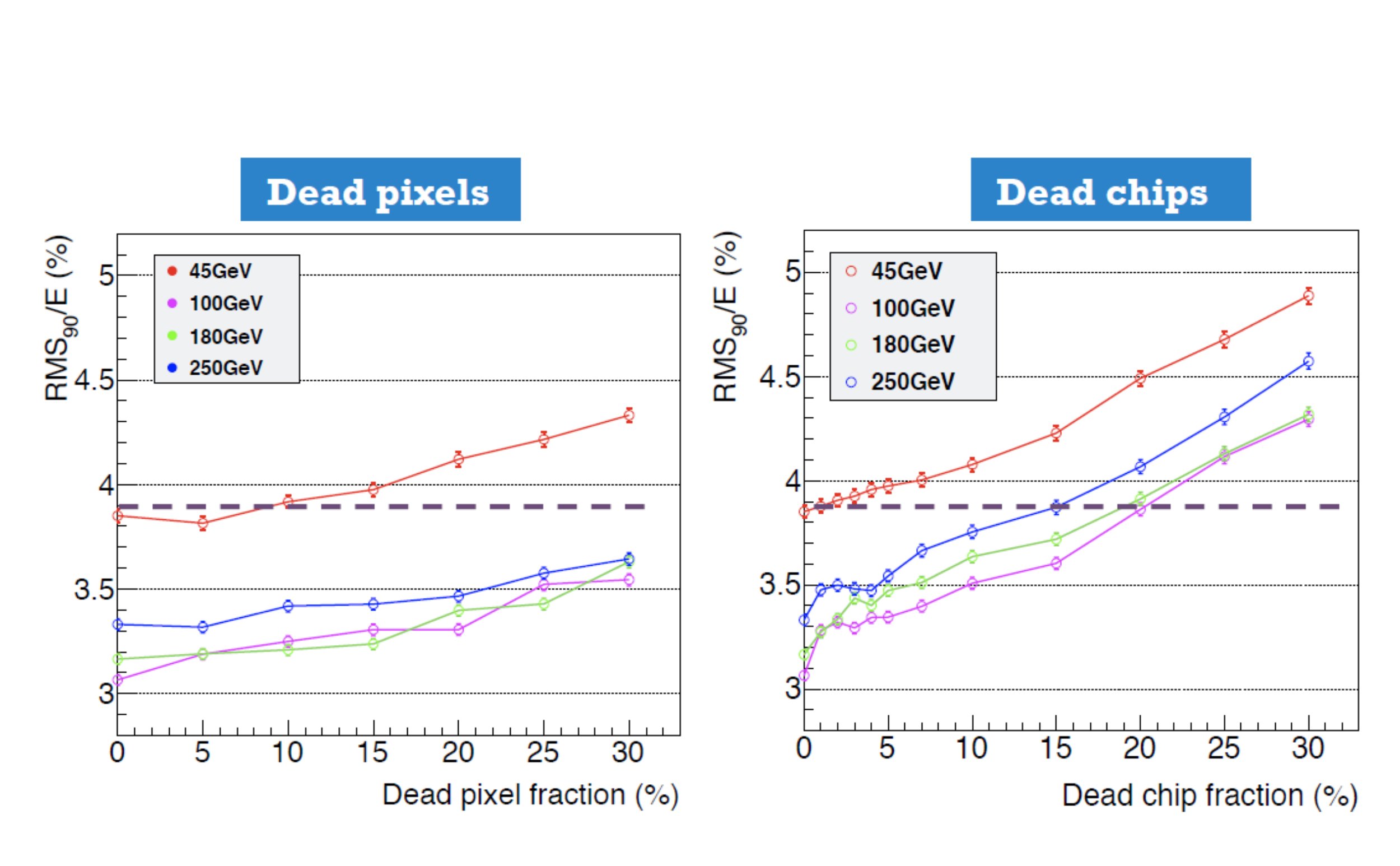}{The dependence of the jet energy resolution on the fraction of dead pixels (left) and readout chips (right).}{deadpixel}

Fig.\ref{deadpixel} shows the dependence of the jet energy resolution on dead pixel and dead chip fractions. 
The jet energy resolution increases for both dead pixels and for dead chips. The jet energy resolution is about two times more 
sensitive to dead readout chips than to dead pixels. 

\subsection{Noisy pixels}

In silicon sensor, typical signal to noise ratio for MIP is about 10. Assuming Gaussian noise, the noise rate over 0.5 MIP threshold is $10^{-7}$. 
Therefore, there are only few tens of noise hits in ECAL. However, in reality, the number of noise hits will be larger due to non-Gaussian noise 
or possibly radiation damage. 
In this study, noisy pixels were distributed uniformly over the detector. Each noise hit was assigned the same energy of 1.4 MIPs, 
an energy somewhat above the threshold, probably larger than the mean of true noise hits.


\afig{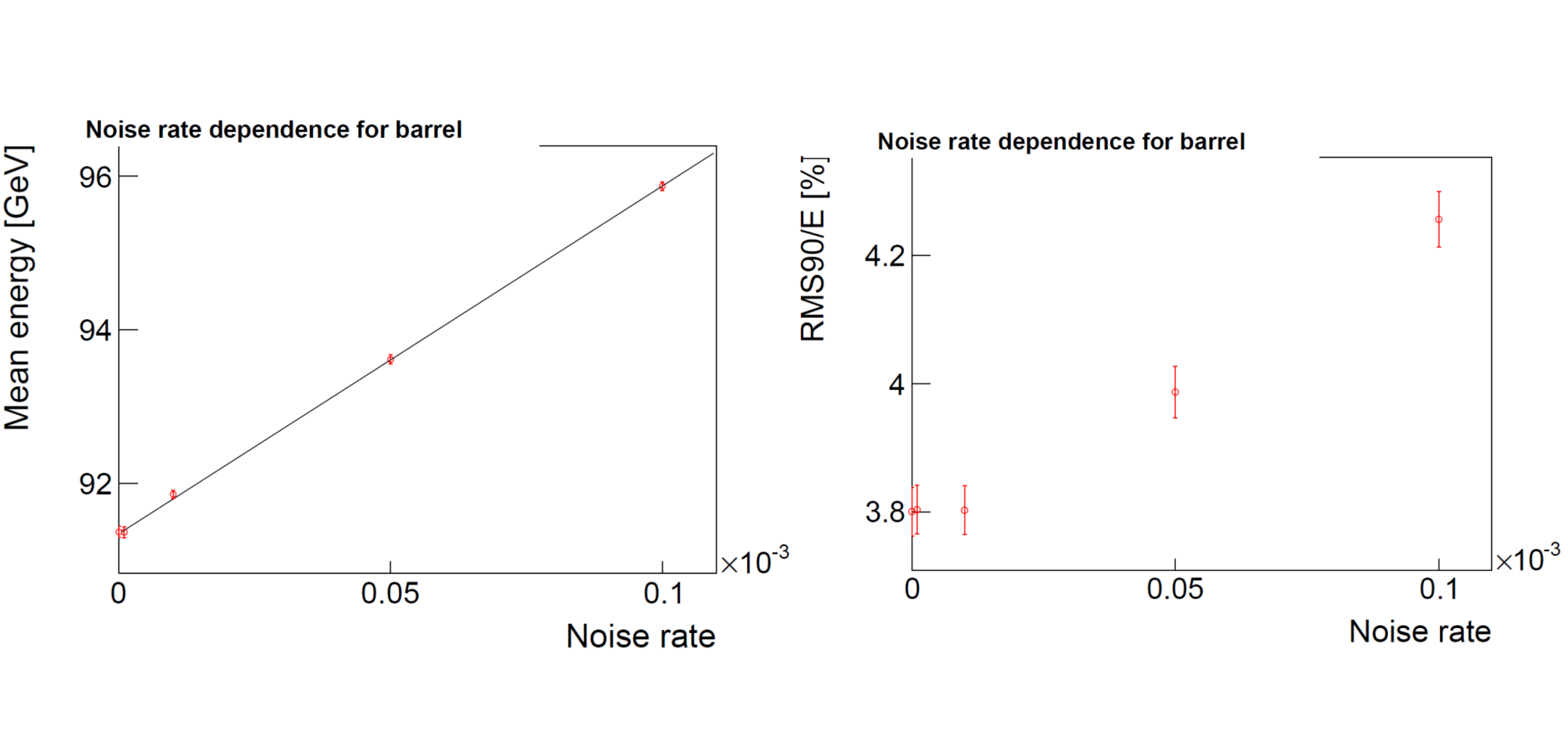}{(left) The dependence of mean measured energy on noise rate. 
(right) The dependence of the jet energy resolution on the noise rate.}{noiseJES}

Noise rates of $10^{-6}$, $10^{-5}$, $5\times 10^{-5}$ and $10^{-4}$ were simulated. Here, the noisy pixels are chosen randomly event by event. 
Fig.\ref{noiseJES} (left) shows the dependence of mean measured energy on the noise rate. The mean measured energy increases linearly, since the
noise was not taken into account in the detector calibration. 
Fig.\ref{noiseJES} (right) shows the dependence of the jet energy resolution on the noise rate. It shows that the jet energy resolution 
is not affected by noise up to a rate of 10$^{-5}$.

\subsection{Mis-calibration}

It is impossible to measure infinitely correct calibration coefficients.
Moreover, there are some possibilities that calibration factors change during running period due to radiation damage, temperature effect, 
electronics problem and so on. In this simulation, calibration coefficients were changed pixel by pixel according to a Gaussian distribution. 
The standard deviation of the Gaussian, $\sigma$ is the parameter of mis-calibration level.

\afig{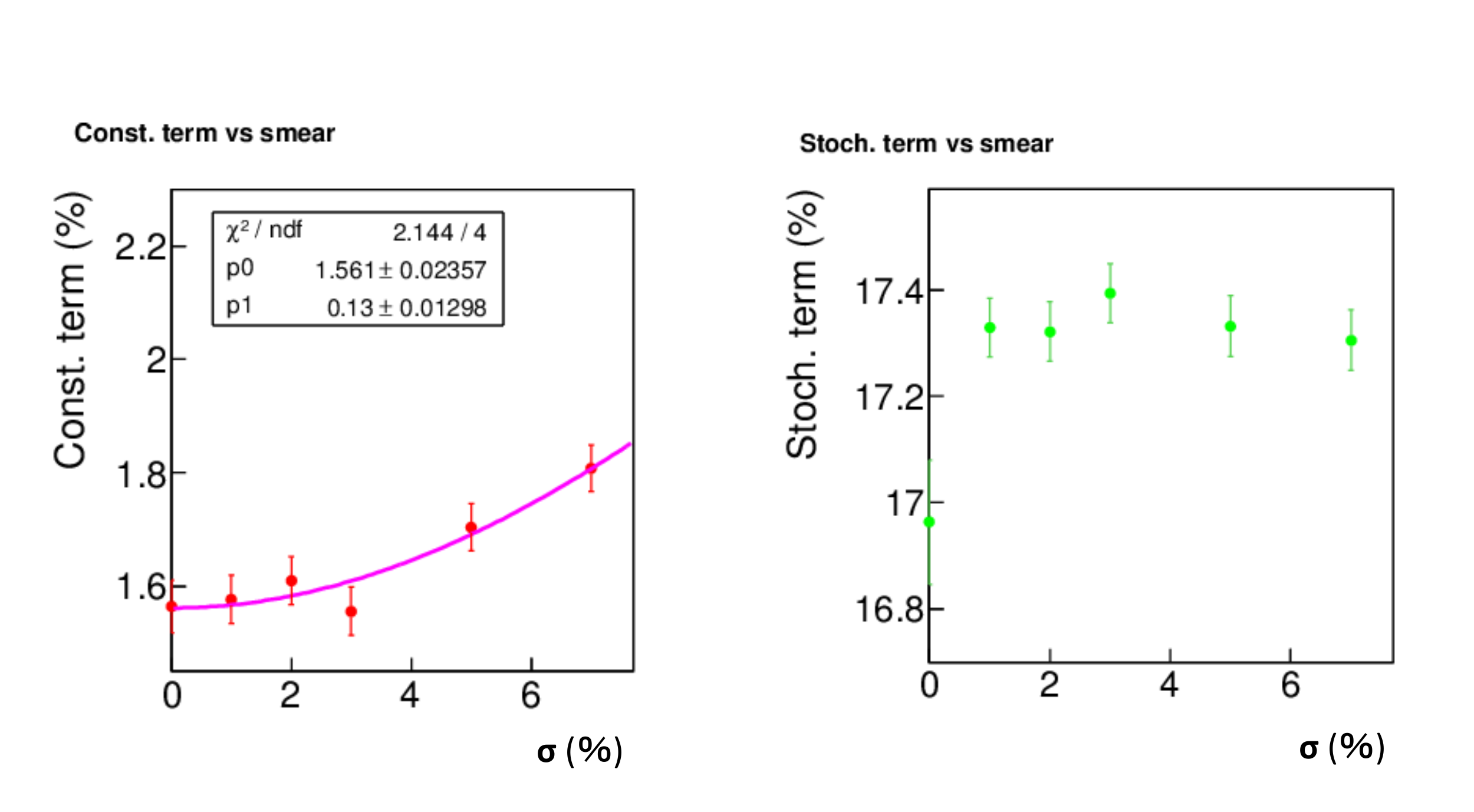}{The photon energy resolution contribution of constant term (left) and stochastic term (right). }{calibconst}
\afig{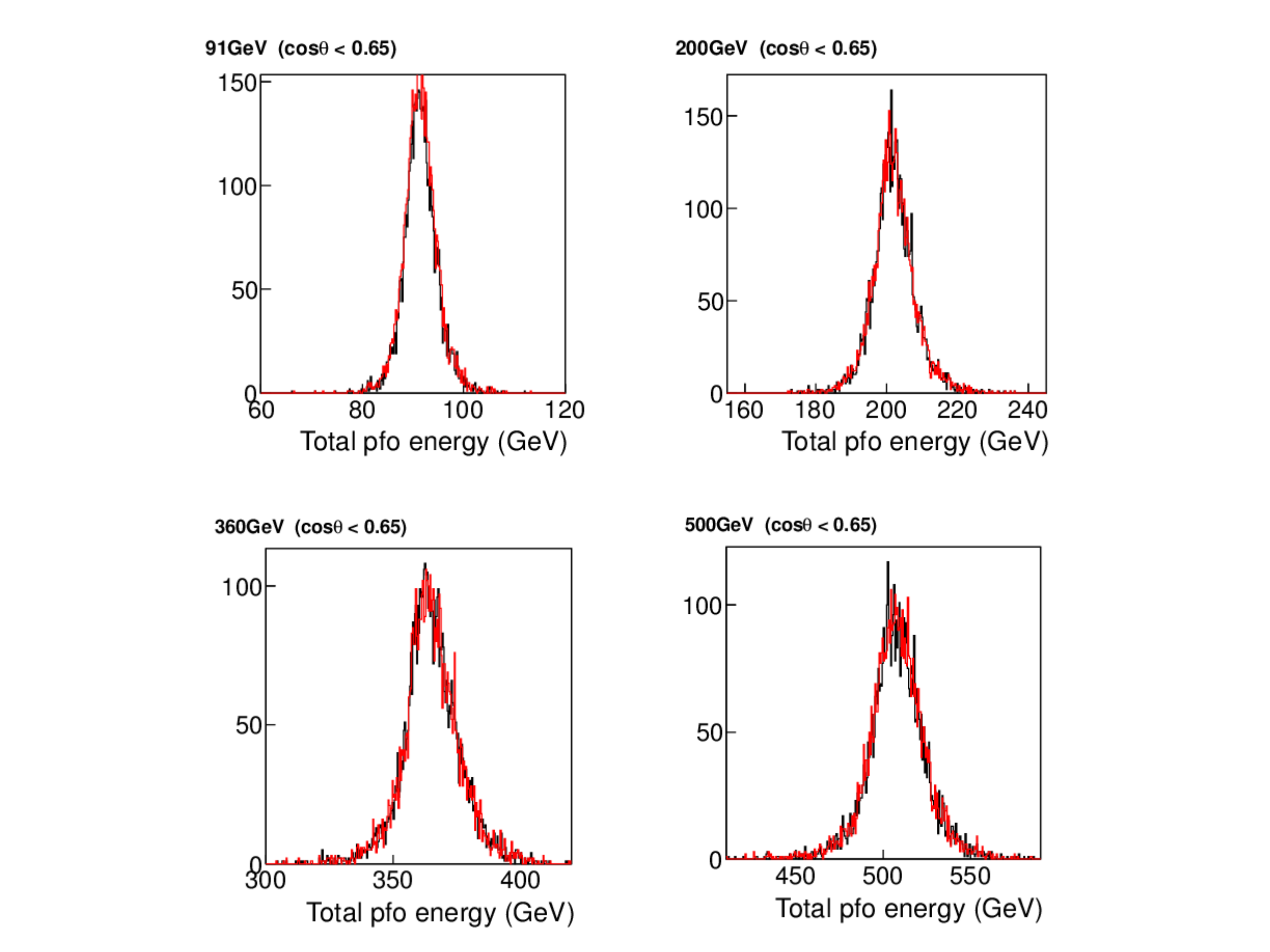}{Histograms of jet energy measurement with correct calibration coefficient (black) and with calibration coefficients
fluctuated by 10\% (red) for several jet energies.}{calibjet}

Simulations were performed for both single photons and jet events. 
Several photon energies were simulated. For each value of $\sigma$, the dependence of the energy 
resolution on energy was fitted by a function of the form:
\begin{equation}
f(E)=\sqrt{b_0^2+\left(\frac{b_1}{\sqrt{E}}\right)^2}.
\end{equation}
Here, $b_0$ is the constant term and $b_1$ the stochastic term. Figure \ref{calibconst} shows the dependence of the constant (left) and 
stochastic (right) terms on $\sigma$. The constant term increases as $\sigma$ increases, while the stochastic term is not affected by $\sigma$. 
This is because the mis-calibration causes non-uniformity, which mainly affects the constant term.

The jet energy measurement simulation results for several energies are shown in Fig.\ref{calibjet}. Here, black histograms show results with 
correct calibration and red histograms show results with mis-calibration of $\sigma=10\%$. There is no significant difference.

\subsection{Cross talk}

Cross talk is caused by readout chip problem, global pedestal shift, or inter-channel cross talk. In this simulation, 
a certain fraction (few \%) of a pixel's energy was added to adjoining pixels.

The simulation was done for photon with several energies. Fig.\ref{crossmean} shows the dependence of the mean measured energy on the cross talk. 
The mean measured energy is normalised by that obtained without cross talk. 
The mean measured energy 
increases approximately linearly
as the cross talk fraction increases. 
The relative increase in reconstructed energy depends on the incident photon energy.

Fig.\ref{crossreso} shows the dependence of the photon energy resolution on the cross-talk. 
The constant term and the stochastic term are derived by same procedure as described in the previous section. 
The constant term is affected by cross talk, while the stochastic term is unchanged.

\afig{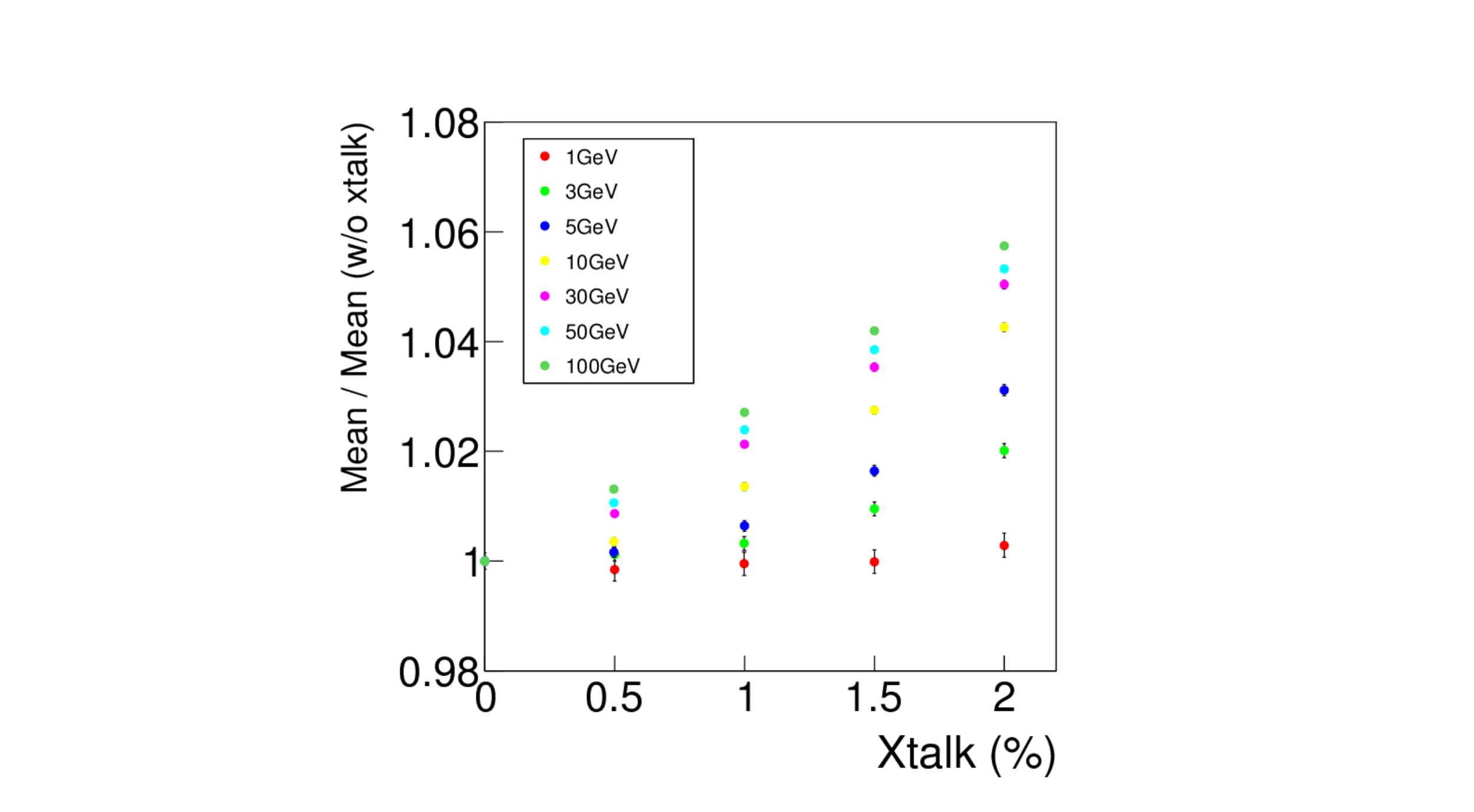}{
The dependence of the constant term of the ECAL energy resolution on the cross-talk.
}{crossmean}
\afig{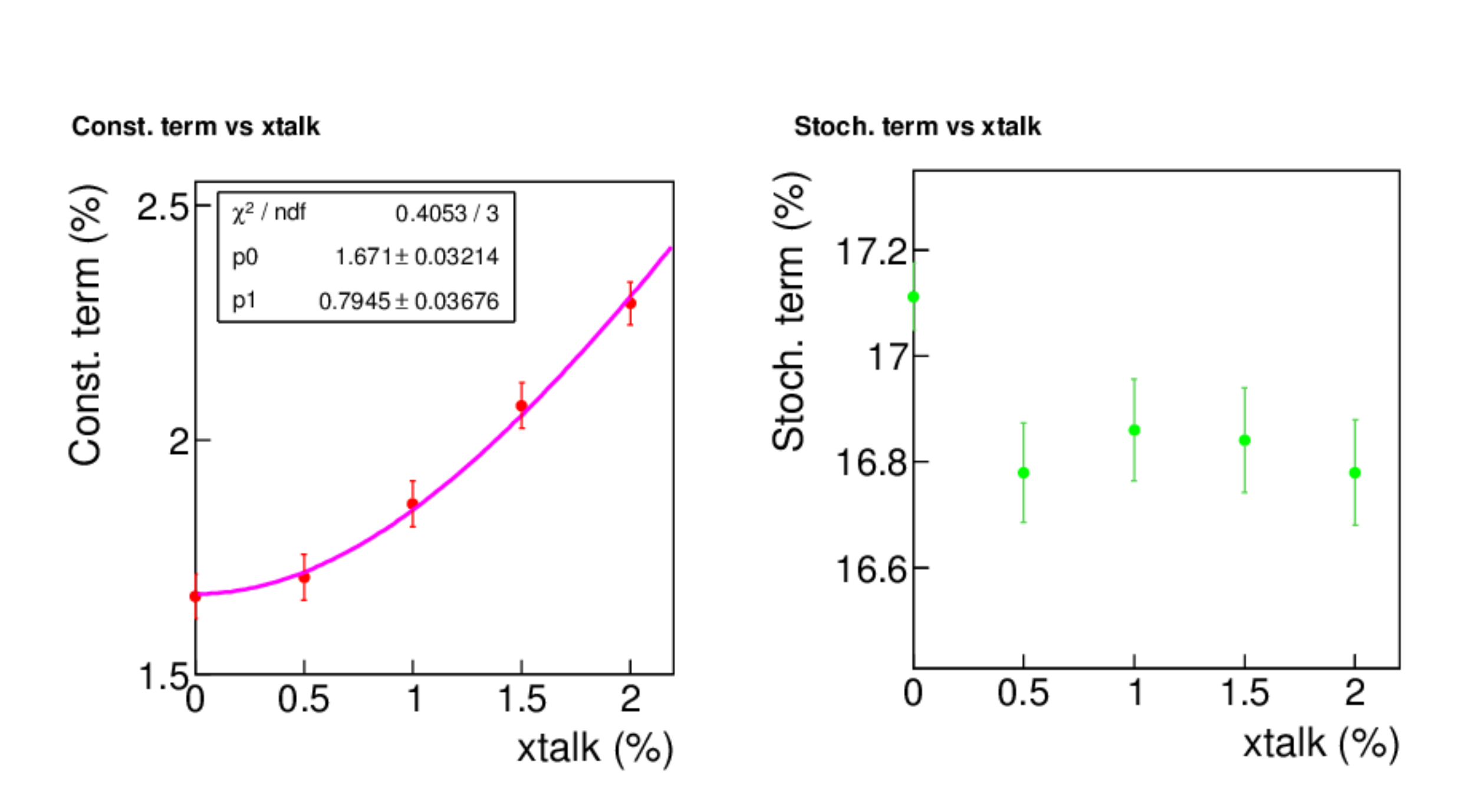}{
The dependence of the stochastic term of the ECAL energy resolution on the cross-talk.
}{crossreso}

\section{Conclusion}

The optimization and robustness of the SiW ECAL for the ILD are being studied.
Jet energy resolution (JER) increases with guard ring width. The relative increase between 0 and 2 mm is about 6 \%.
With increasing PCB thickness, JER starts to degrade at around 3 mm. Between 3 mm and 5mm, the increase in JER is 5-8 \%.
The JER is rather insensitive to random dead channels in the ECAL; fractions of up to 10\% have relatively little effect.
JER is more sensitive to dead readout chips, about 2 times more than for dead pixels. 
Noise rates less than $10^{-5}$ may be tolerable (for current implementations of PFA).
Mis-calibration and inter-pixel cross talk affect the constant term of the photon energy resolution.


\begin{thebibliography}{99}
	\bibitem{TDR1}Ties Behnke et al. (ed.s), "The International Linear Collider Technical Design Report Volume 1. Executive summary", arXiv.1306.6327, (2013)
	\bibitem{DBD}Ties Behnke et al. (ed.s), "The International Linear Collider Technical Design Report Volume 4. Detectors", arXiv.1306.6329, (2013)
	\bibitem{PFA1}
J.-C. Brient and H. Videau, ``The calorimetry at the future e + e − linear collider'', hep-ex/0202004.
	\bibitem{PFA2}
M. Thomson, "Particle flow calorimetry and the PandoraPFA Algorithm", Nucl.Instrum.Meth.A611:25-40, (2009)
	\bibitem{Mokka}P. de Freitas, et al., http://polzope.in2p3.fr:8081/MOKKA.
	\bibitem{GEANT4}GEANT4 collaboration, S. Agostinelli et al., Nucl. Instr. and Meth. A506 (2003) 3;
GEANT4 collaboration, J. Allison et al., IEEE Trans. Nucl. Sci. 53 (2006) 1.
	\bibitem{Marlin}F. Gaede, Nucl. Instr. and Meth. A559 (2006) 177.

\end{thebibliography}
\end{document}